\documentclass[twoside]{amsart}
\usepackage[letterpaper,hmargin={1.25in,1.25in},top=1.25in,bottom=1.25in]{geometry}

\usepackage{amsmath,amssymb,amsthm}
\usepackage{setspace}
\usepackage{graphicx}

\usepackage[foot]{amsaddr}
\usepackage{natbib}

\newtheorem{lemma}{Lemma}
\newtheorem{theorem}{Theorem}

\usepackage[small]{caption}

\usepackage{algorithm}

\newcommand{\R}{\mathbb{R}}
\newcommand{\makeright}[2]{\ifx#1\left\right#2\else#1#2\fi}
\newcommand{\ERW}[2][\left]{\mathbf{E}    #1[ #2 \makeright{#1}{]}}
\newcommand{\PR}[2][\left]{\mathrm{P}    #1[ #2 \makeright{#1}{]}}
\newcommand{\VAR}[2][\left]{{\rm Var}  #1[ #2 \makeright{#1}{]}}

\newcommand{\N}{\mathcal{N}}
\newcommand{\ESS}{\textsc{ess}}
\newcommand{\DIV}{\textsc{div}}
\newcommand{\RMSE}{\textsc{rmse}}
\newcommand{\CRPS}{\textsc{crps}}
\newcommand{\EnKF}{ensemble Kalman filter}
\newcommand{\EnKPF}{ensemble Kalman particle filter}

\begin{document}

\onehalfspacing
\title{Bridging the ensemble Kalman and particle filter}

\author{Marco Frei}
\address{$^1$Seminar for Statistics, ETH Zurich, CH-8092 Zurich, Switzerland}
\email{frei@stat.math.ethz.ch}

\author{Hans R. K\"unsch$^{1}$}
\email{kuensch@stat.math.ethz.ch}

\date{July 31, 2012}

\begin{abstract}
  In many applications of Monte Carlo nonlinear filtering, the propagation
  step is computationally expensive, and hence, the sample size is
  limited. With small sample sizes, the update step becomes crucial.
  Particle filtering suffers from the well-known problem of sample
  degeneracy. Ensemble Kalman filtering avoids this, at the expense of
  treating non-Gaussian features of the forecast distribution
  incorrectly. Here we introduce a procedure which makes a continuous
  transition indexed by $\gamma \in [0,1]$ between the ensemble and the
  particle filter update. We propose automatic choices of the parameter
  $\gamma$ such that the update stays as close as possible to the particle
  filter update subject to avoiding degeneracy. In various examples, we
  show that this procedure leads to updates which are able to handle
  non-Gaussian features of the prediction sample even in high-dimensional
  situations.
\end{abstract}

\maketitle

\markboth{\MakeUppercase{M. Frei \and H. R. K\"unsch}}
{\uppercase{Bridging the ensemble Kalman and particle filter}}

\section{Introduction}
State space models consist of a (discrete or continuous time) Markov
process which is partially observed at discrete time points and subject to
independent random errors.  Estimation of the state at time $t$ given
observations up to the same time is called filtering or data
assimilation. Since exact computations are possible essentially only in
linear Gaussian situations, mostly Monte Carlo methods are used for
filtering. In many environmental applications, in particular in atmospheric
physics, oceanography and reservoir modelling, the dimension of the state
is, however, very large and the computational costs to propagate the state
forward in time are huge, which severely limits the potential sample size
for Monte Carlo filtering methods. Particle filters
\citep{gordon93,pitt99,doucet00} suffer from the well-known problem of
sample degeneracy \citep{snyder08}. In contrast, the ensemble Kalman filter
\citep{evensen94,burgers98,houtekamer98} can handle some problems where the
dimensions of states and observations are large, and the number of
replicates is small, but at the expense of incorrectly treating
non-Gaussian features of the forecast distribution that arise in nonlinear
systems.

To relax the Gaussian assumption, two paradigms are predominant: mixture
filters that approximate the forecast distribution as a mixture of
Gaussians \citep{bengtsson03,sun09,dovera11,frei11,hoteit12,rezaie12}, and
sequential importance samplers that use the \EnKF{} as a proposal distribution
\citep{mandel09,papadakis10}. In this article, we introduce an update
scheme that blends these two flavours: A Gaussian mixture proposal obtained
from an \EnKF{} update based on a tempered likelihood is corrected by a
particle filter update. In this way we do not have to fit a Gaussian
mixture to the forecast sample nor do we have to approximate the ratio of
the predictive density to the proposal density. A further advantage of our
procedure is that we can implement these two steps in such a way that the
particle weights do not depend on artificial observation noise variables
and the resampling avoids ties. 

Our procedure depends on a single tuning parameter
$\gamma \in [0,1]$, which allows continuous interpolation between the \EnKF{}
($\gamma=1$) and the particle filter ($\gamma=0$). Hence, the parameter
$\gamma$ controls the bias-variance trade-off between a correct update and
maintaining the diversity of the sample. It can be chosen without prior
knowledge based on a suitable measure of diversity like effective sample
size \ESS \, \citep{liu96}, or the expected number of Gaussian components 
which are represented in the resample. 

The rest of the article is organized as follows. In
Section~\ref{sec:notation}, we detail the problem setting, introduce some
notation, and provide background material. In Section~\ref{sec:EnKPF}, we
present our new method and discuss implementational aspects. In
Section~\ref{sec:choiceofgamma}, we discuss the choice of the tuning
parameter $\gamma$. In Section~\ref{sec:ex-single}, we consider numerical
examples that involve single updates only; based on different prior
specifications, we examine the differences of our method in comparison to
the \EnKF{} and the particle filter. In Section~\ref{sec:ex-several}, we
consider examples that involve many update cycles in two common test
beds. Section~\ref{sec:outlook} contains an outlook
to possible generalizations.

\section{Problem setting, notation and background material}
\label{sec:notation}
We consider a dynamical system with state variable $(x_t \in \R^q: t=0,1,
\ldots)$ and observations $(y_t \in \R^r: t=1,2, \ldots)$. The state
follows a deterministic or stochastic Markovian evolution, that is
$x_t=g(x_{t-1}, \xi_t)$ where the system noise $\xi_t$ is independent of
all past values $x_s$ and all $\xi_s$, $s<t$. There is no need to know the
function $g$ in explicit form, we only assume that for given $x_{t-1}$ we
are able to simulate from the distribution of $g(x_{t-1}, \xi_t)$. In
particular, the evolution can be in continuous time, given by an ordinary
or stochastic differential equation.

In all cases we assume linear observations with Gaussian noise: $y_t = H
x_t + \epsilon_t, \quad \epsilon_t \sim \N(0,R)$.  This means that the
likelihood for the state $x_t$ given the observation $y_t$ is
$\ell(x_t|y_t) = \varphi(y_t; Hx_t, R)$.  Here and in the following
$\varphi(x; \mu, \Sigma)$ denotes the (in general multivariate) normal
density with mean $\mu$ and covariance $\Sigma$ at $x$. In the final
section, we will discuss briefly how to adjust the method for non-Gaussian
likelihoods.

We denote all observations up to time $t$, $(y_1, \ldots, y_t)$ by
$y_{1:t}$. The forecast distribution $\pi^p_t$ at time $t$ is the conditional
distribution of $x_t$ given $y_{1:t-1}$, and the filter distribution
$\pi_t^u$ at time $t$ is the conditional distribution of $x_t$ given
$y_{1:t}$. In principle, these distributions can be computed recursively,
alternating between propagation and update steps. The propagation step
leads from $\pi_{t-1}^u$ to $\pi_t^p$: $\pi^p_t$ is the 
distribution of $g(x_{t-1},\xi_t)$
where $x_{t-1} \sim \pi^u_{t-1}$ and $\xi_t$ is independent of $x_{t-1}$
and has the distribution given by the evolution of the system.
The update step leads from $\pi_{t}^p$ to $\pi_t^u$
and is nothing else than Bayes formula:
$\pi_t^u(dx_t) \propto \ell(x_t|y_t) \pi_t^p(dx_t)$. However,
analytical computations are possible (essentially) only if the system
evolution is also linear with additive Gaussian noise. Hence one resorts to
Monte Carlo approximations, i.e., one represents $\pi_t^p$ and $\pi_{t}^u$
by ensembles (samples) $(x^p_{t,j})$ and $(x^u_{t,j})$ respectively. The
members of these ensembles are called particles.

The propagation step just lets the particles evolve according to the
dynamics of the state, that is we simulate according to the time evolution
starting at $x^u_{t-1,j}$ at time $t-1$, $x^p_{t,j} = g(x^u_{t-1,j},
\xi_{t,j})$.  However, the computational complexity of this step limits the
number of particles, that is the size of the sample.

The (bootstrap) particle filter \citep{gordon93} updates the forecast
particles by weighting with weights proportional to the likelihood
$\ell(x_t|y_t)$ and converts this into an unweighted sample by resampling,
i.e., $(x^u_{t,j})$ is obtained by sampling from
\begin{equation}
\label{eq:pf-mixt}
\sum_{j=1}^N \omega_{t,j} \Delta_{x^p_{t,j}}, \quad \omega_{t,j} =
\frac{\ell(x^p_{t,j}|y_t)}{\sum_{k=1}^N \ell(x^p_{t,k}|y_t)}.
\end{equation}
Thus some of the forecast particles disappear and others are replicated.
If the likelihood is quite peaked, which is the case in high dimensions
with many independent observations, the weights will be heavily unbalanced,
and the filter sample eventually degenerates since it concentrates on a
single or a few particles; see \cite{snyder08}.  Auxiliary
particle filters \citep{pitt99} can attenuate this behaviour to some
extent, but they require good proposal distributions for the propagation,
and an analytical expression for the transition densities.

The ensemble Kalman filter \citep{burgers98, houtekamer98} makes an
affine correction of the forecast particles based on the new observation
$y_t$ and artificial observation noise variables $\epsilon_{t,j} \sim
\N(0,R)$:
\[ x^{u}_{t,j} = x^p_{t,j} + K(\hat{P}_t^p)(y_t -H x^p_{t,j} +
\epsilon_{t,j}) \] where $\hat{P}_t^p$ is an estimate of the forecast
covariance at time $t$, typically a regularized version of the sample
covariance of $(x^p_{t,j})$, and $K(P)$ is the Kalman gain $K(P) = P
H'(HPH' +R)^{-1}$. This update formula is (asymptotically) correct under
the assumption that the forecast distribution $\pi^p_t$ is Gaussian
\citep{legland10}. Although this is usually not valid, the update
nevertheless has been found to work well in a variety of situations; see
\cite{evensen07} and references therein. For later use, we note that
conditional on the forecast sample,
\[ x^{u}_{t,j} \sim \N(x^p_{t,j} + K(\hat{P}_t^p)(y_t -H x^p_{t,j}), 
K(\hat{P}_t^p) R K(\hat{P}_t^p)'). \]
Therefore the filter sample can be considered as a balanced sample from the
(conditional) Gaussian mixture
\begin{equation}
\label{eq:bal-mixt}
\frac{1}{N} \sum_{j=1}^N \N(x^p_{t,j} + K(\hat{P}_t^p)(y_t -H x^p_{t,j}), 
K(\hat{P}_t^p) R K(\hat{P}_t^p)').
\end{equation}
Here, ``balanced sample'' simply means that we draw exactly one realization
from each of the $N$ equally weighted Gaussian components.

\section{A bridge between ensemble and particle updates: The ensemble
  Kalman particle filter}
\label{sec:EnKPF}
\subsection{The new method}
We consider here the update at a single fixed time $t$ and thus suppress
$t$ in the notation.  We follow the ``progressive correction'' idea
\citep{musso01} and write
$$\pi^u(dx) \propto \pi^{u,\gamma}(dx) \ell(x|y)^{1-\gamma}, \quad
\pi^{u,\gamma}(dx) \propto \pi^{p}(dx) \ell(x|y)^\gamma$$ where $0 \leq
\gamma \leq 1$ is arbitrary. Our approach is to use an ensemble Kalman
filter update to go from $\pi^{p}$ to $\pi^{u,\gamma}$, and a particle
filter update to go from $\pi^{u,\gamma}$ to $\pi^u$.  The rationale behind
this two-stage procedure is to achieve a compromise between sample
diversity and systematic error due to non-Gaussian features of $\pi^p$. The
former is large if $\gamma$ is close to one because the \EnKF{} update
draws the particles closer to the observation $y$, and the exponent
$1-\gamma$ dampens the ratio of any two resampling probabilities.  The
latter is small if $\gamma$ is small.

Since $\ell(x|y)^\gamma \propto \varphi(y; Hx, R/ \gamma)$, and
\begin{equation}
\label{gain-gammaP}
PH'(HPH' + R/\gamma)^{-1} = \gamma P H'(\gamma HPH' +R)^{-1} = K(\gamma P)
\end{equation}the
ensemble filter update is straightforward: We only need to compute the gain
with the reduced covariance $\gamma \hat{P}^p$. The particle update then
resamples with weights proportional to $\ell(x|y)^{1-\gamma} \propto
\varphi(y; Hx, R/(1-\gamma))$.  
However, there are two immediate drawbacks to such an algorithm:
the particle weights depend on the artificial observation
noise variables which are needed for the \EnKF{} update, and the resampling
introduces tied values. We show next how to address both points. By
\eqref{eq:bal-mixt}, we can write
\begin{equation}
  \label{update-step1}
\pi^{u,\gamma} \approx \pi^{u,\gamma}_{\rm{EnKF}} =\frac{1}{N} \sum_{j=1}^N \mathcal{N}(\nu^{u,\gamma}_j, 
Q(\gamma,\hat{P}^p))
\end{equation}
where 
\begin{align}
\label{eq:nu}
 \nu^{u,\gamma}_j &= x^p_j + K(\gamma \hat{P}^p)(y - H x^p_j),\\
\label{eq:Q}
 Q(\gamma,\hat{P}^p) &= \frac{1}{\gamma} K(\gamma \hat{P}^p) 
R K(\gamma \hat{P}^p)'.
\end{align}
Instead of sampling from \eqref{update-step1} and applying a particle
correction, we delay the \EnKF{} sampling step, and update
\eqref{update-step1} analytically. This is easy because the update of a
Gaussian mixture by a Gaussian likelihood is again a Gaussian mixture whose
parameters can be computed easily \citep{alspach72}. We obtain
\begin{equation}
  \label{update-step2}
\pi^u \approx \pi^{u}_{\rm{EnKPF}} = \sum_{j=1}^N \alpha^{u,\gamma}_j \mathcal{N}(\mu^{u,\gamma}_j, P^{u,\gamma})  
\end{equation}
where EnKPF stands for ensemble Kalman
particle filter and 
\begin{align}
\alpha^{u,\gamma}_j &\propto \varphi(y;H \nu^{u,\gamma}_j,H Q(\gamma,\hat{P}^p)H' + 
\frac{1}{1-\gamma}R), \label{eq:alpha}\\
\mu^{u,\gamma}_j &= \nu^{u,\gamma}_j + K((1-\gamma)Q(\gamma,\hat{P}^p))
(y - H \nu^{u,\gamma}_j), \label{eq:mu}\\
P^{u,\gamma} &= (I - K((1-\gamma)Q(\gamma,\hat{P}^p))H)Q(\gamma,\hat{P}^p).
\label{eq:Pu}
\end{align}
The update consists now in sampling from \eqref{update-step2}. The mixture
proportions $\alpha^{u,\gamma}_j$ do not depend on the artificial
observation noise variables, and even if one $\alpha^{u,\gamma}_j$
dominates, there is still some diversity in the filter sample because the
covariance $P^{u,\gamma}$ is not zero if $\gamma>0$.

Sampling from the $j$-th component of (\ref{update-step2}) can be done 
as follows: Let $\epsilon_1$ and $\epsilon_2$ be two independent 
$\mathcal{N}(0,R)$ random variables. Then
\begin{equation}
\label{br-update-1}x^{u,\gamma}=x^p_j + K(\gamma \hat{P}^p)
(y + \frac{\epsilon_{1}}{\sqrt{\gamma}}-Hx^p_j) = \nu^{u,\gamma}_j + 
K(\gamma \hat{P}^p) \frac{\epsilon_{1}}{\sqrt{\gamma}}
\end{equation}
clearly has distribution $\mathcal{N}(\nu^{u,\gamma}_j,
Q(\gamma,\hat{P}^p))$, and thus by standard arguments
\begin{equation*}
x^u = x^{u,\gamma} + K((1-\gamma)Q(\gamma,\hat{P}^p))
\left(y + \frac{\epsilon_{2}}{\sqrt{1-\gamma}}- H x^{u,\gamma}\right)
\end{equation*}
is a sample from $\mathcal{N}(\mu^{u,\gamma}_j, P^{u,\gamma})$.
Hence there is no need to compute a square root
of $P^{u,\gamma}$. 

To summarize, given a forecast ensemble $(x^p_j)$ and an observation $y$,
the ensemble Kalman particle filter consists of the following steps:

\begin{algorithm}[h]
\caption{Ensemble Kalman particle filter} \label{al1}
\begin{tabbing}
  1. Compute the estimated forecast covariance $\hat{P}^p$. \\
  2. Choose $\gamma$ and compute $K(\gamma \hat{P}^p)$ according to
  \eqref{gain-gammaP} and 
  $\nu^{u,\gamma}_j$ according to \eqref{eq:nu}. \\
  3. Compute $Q(\gamma,\hat{P}^p)$ according to
  \eqref{eq:Q} and the weights $\alpha^{u,\gamma}_j$ according to
  \eqref{eq:alpha}. \\
  4. Choose indices $I(j)$ by sampling from the weights
  $\alpha^{u,\gamma}_j$ with some \\ \qquad balanced
 sampling scheme; e.g., equation (12) in \cite{kuensch05}. \\
  5. Generate $\epsilon_{1,j} \sim \N(0,R)$ and set
$x^{u,\gamma}_j = \nu^{u,\gamma}_{I(j)} + 
K(\gamma \hat{P}^p) \frac{\epsilon_{1,j}}{\sqrt{\gamma}}$. \\
  6. Compute $K((1-\gamma)Q(\gamma,\hat{P}^p))$, generate $\epsilon_{2,j}
  \sim \N(0,R)$ and set \\
\qquad $x^u_j = x^{u,\gamma}_j + K((1-\gamma)Q(\gamma,\hat{P}^p))
\left(y + \frac{\epsilon_{2,j}}{\sqrt{1-\gamma}}- H x^{u,\gamma}_j
\right)$.
\end{tabbing}
\end{algorithm}


Because matrix inversion is continuous, it is easy to check that as
$\gamma \rightarrow 0$, $\nu^{u,\gamma}_j \rightarrow x^p_j$,
$Q(\gamma,\hat{P}^p) \rightarrow 0$, 
$\alpha^{u,\gamma}_j \rightarrow \varphi(y;Hx^p_j,R)$,
$\mu^{u,\gamma}_j \rightarrow x^p_j$ and $P^{u,\gamma} \rightarrow 0$.
Hence in the limit $\gamma \rightarrow 0$, we obtain the particle filter
update. Similarly, in the limit $\gamma \rightarrow 1$ we obtain the 
ensemble Kalman filter update because for $\gamma \rightarrow 1$
$(H Q(\gamma,\hat{P}^p)H' + R/(1-\gamma))^{-1}$ converges to zero
and thus $\alpha^{u,\gamma}_j \rightarrow 1/N$. The \EnKPF{} therefore provides
a continuous interpolation between the particle and the ensemble Kalman
filter. 

\subsection{Modifications in high dimensions}
\label{high-dim}
If the dimension $q$ of the state space is larger than the number $N$ of
particles, then the variability of the usual sample covariance is huge, and
one should use a regularized version as estimate $\hat{P}^p$. In the
context of the \EnKF{}, the standard regularization technique is the use of a
tapered estimate, that is we multiply the sample covariance matrix by a
correlation matrix which is zero as soon as the distance between the two
components of the state is larger than some threshold, see e.g.,
\cite{houtekamer01, furrer07}. If also the error covariance matrix $R$ and
the observation matrix $H$ are sparse, then this tapering has the
additional benefit that the computation of $\nu^{u,\gamma}_j$ in step~2 and
of $x^{u,\gamma}_j$ in step~5 is much faster because we do not need to
compute $K(\gamma \hat{P}^p)$ for this. It is sufficient to solve $2N$
equations of the form $(\gamma HPH' +R)x=b$ which is fast for sparse
matrices. However, this advantage is lost because for $Q(\gamma, \hat{P}^p)$
we need $K(\gamma \hat{P}^p)$, which is in general a full matrix. We could
multiply the gain matrix by another taper in order to facilitate the
computation of $Q(\gamma, \hat{P}^p)$ and to turn $Q(\gamma, \hat{P}^p)$ into
a sparse matrix.  This would then make steps~4 and~6 in the algorithm above
faster because again all we need to do is to solve $2N$ equations of the
form $((1-\gamma)HQ(\gamma,\hat{P}^p)H'+R)x=b$. Alternatively, one can
could also replace the gain by the optimal matrix with a given sparsity
pattern, i.e., using a localized update in grid space, see \cite{sakov10}.

Because $K(\gamma \hat{P}^p)$ is only used to compute $Q(\gamma,
\hat{P}^p)$, a simpler alternative which avoids computing gain matrices is
to generate the values $K(\gamma \hat{P}^p)
\epsilon_{1,j}/\sqrt{\gamma}$ needed in step~5 before step~3 and~4, and
then to replace $Q(\gamma, \hat{P}^p)$ by a sparse regularized version of
the sample covariance of these values. If this approach is taken, it is
usually feasible to generate more than $N$ such values in order to reduce
the Monte Carlo error in the regularized sample covariance matrix.

\subsection{Consistency of the \EnKPF{} in the Gaussian case}
We establish consistency of the \EnKPF{} for any $\gamma$ as
the ensemble size $N$ tends to infinity, provided that the forecast sample
is iid normal. We assume that all random quantities are defined on some
given probability space $(\Omega,\mathcal{F},\mathrm{P})$ and ``almost
surely'' is short for ``$\mathrm{P}$-almost surely''. The observation $y$
is considered to be fixed (nonrandom). The superscript $\cdot^N$ is added
to any random quantity that depends on the ensemble size $N$. We use the
following notion of convergence: A sequence $(\pi^N)_{N \in \mathbb{N}}$ of
random probability measures converges almost surely weakly to the
probability measure $\pi$ if $\int h(x) \pi^N(dx)$ converges almost surely
to $\int h(x) \pi(dx)$ for any continuous and bounded function $h$.

\begin{theorem}
\label{thm:consistency}
Suppose that $(x^p_j)_{j \in \mathbb{N}}$ is an iid sample from
$\pi^p=\N(\mu^p,P^p)$. Then, for any $\gamma \in [0,1]$, the sequence
$\pi^{u,N}_{\rm{EnKPF}}$ as defined in \eqref{update-step2} converges
almost surely weakly to the true posterior $\pi^u(dx) \propto \varphi(y;Hx,R)
\pi^p(dx)$. Additionally, if $(x^{u,N}_j)_{j=1,\dots,N}$ is a conditionally
iid sample from $\pi^{u,N}_{\rm{EnKPF}}$, then also
$\frac{1}{N}\sum_{j=1}^N \Delta_{x^{u,N}_j}$ converges almost surely weakly
to $\pi^u$.
\end{theorem}
A proof is given in the appendix. Notice that if balanced sampling is used
to sample from $\pi^{u}_{\rm{EnKPF}}$, the particles are no longer
conditionally iid, and the arguments become more complicated; see
\cite{kuensch05} for a discussion in the context of auxiliary particle
filters. Inspection of the proof of Theorem~\ref{thm:consistency} shows
that if $\pi^p$ is non-Gaussian with finite second moments, then
$\pi^{u}_{\rm{EnKPF}}$ still converges almost surely weakly to a nonrandom
limit distribution $\pi^{u,\infty}_{\rm{EnKPF}}$. The limit distribution
depends on $\gamma$ and generally differs from the correct posterior
$\pi^u$ for $\gamma>0$. The limit cannot be easily identified, and in
particular it is difficult to quantify the systematic error as a function of
$\gamma$. Using similar arguments as in \cite{randles82}, it is also
possible to show that
\[ N^{1/2}\left(\int h(x) \pi^{u,N}_{\rm{EnKPF}}(dx) - \int h(x)
  \pi^{u,\infty}_{\rm{EnKPF}}(dx)\right) \to \N(0,V)\] weakly, where the
asymptotic covariance $V$ depends on $h$, $\pi^p$ and $\gamma$. In general,
$V$ is analytically intractable, and we cannot verify if $V$ decreases as a
function of $\gamma$, as we expect.

\section{Choice of $\gamma$}
\label{sec:choiceofgamma}

\subsection{Asymptotics of weights}
Recall that for $\gamma=0$ the method is exactly the particle filter, and
for $\gamma=1$ it is exactly the \EnKF{}. Hence it is clear that there is a
range of values $\gamma$ where we obtain an interesting compromise between
the two methods in the sense that the weights $(\alpha^{u,\gamma}_j)$ are
neither uniform nor degenerate. We try to provide some theoretical insight
where this range of values $\gamma$ is, and later we develop a criterion
which chooses a good value $\gamma$ automatically.

We want to see how the weights $\alpha^{u,\gamma}_j$ in (\ref{update-step2}) 
behave as a function of $\gamma$ when the dimension of the 
observations is large. By definition
\begin{align*}
\alpha^{u,\gamma}_j &\propto \exp(-\frac{1}{2}(y - H\nu^{u,\gamma}_j)' 
(H Q(\gamma,\hat{P}^p)H' + \frac{1}{1-\gamma}R)^{-1} (y - H\nu^{u,\gamma}_j))\\
&\propto \exp(-\frac{1}{2}(x^p_j -\mu^p)' \hat{C}_\gamma (x^p_j -\mu^p) 
+ \hat{d}_\gamma'(x^p_j -\mu^p))
\end{align*}
where $\mu^p$ is the prediction mean,
\begin{align*} 
\hat{C}_\gamma&=(1-\gamma) H'(I-\hat{K}_\gamma'H')((1-\gamma) H
\hat{Q}_\gamma H' + R)^{-1} (I-H\hat{K}_\gamma) H, \\
\hat{d}_\gamma&=(1-\gamma) H'(I-\hat{K}_\gamma'H')((1-\gamma) H 
\hat{Q}_\gamma H' + R)^{-1} (I-H\hat{K}_\gamma)(y-H \mu^p)
\end{align*}
and $\hat{K}_\gamma$ and $\hat{Q}_\gamma$ stand for $K(\gamma \hat{P}^p)$ and 
$Q(\gamma, \hat{P}^p)$.

The following lemma gives an approximate formula for the
variance of the $\alpha^{u,\gamma}_j$.

\begin{lemma}
\label{lemma:weights}
Define approximate weights by 
$$\tilde{\alpha}^{u,\gamma}_j = \frac{1}{N} \frac{\exp(-\frac{1}{2}(x^p_j
  -\mu^p)' C_\gamma
  (x^p_j -\mu^p) + d_\gamma'(x^p_j -\mu^p))}{\ERW{\exp(-\frac{1}{2}(x^p_j
    -\mu^p)' C_\gamma (x^p_j -\mu^p) + d_\gamma'(x^p_j -\mu^p))}}.$$
where $C_\gamma$ and $d_\gamma$ are as defined above, but with the true forecast
covariance $P^p$ instead of $\hat{P}^p$. If the forecast sample is
iid $\N(\mu^p,P^p)$, then 
$$N^2 \VAR{\tilde{\alpha}^{u,\gamma}_j} = \frac{\det(P^p C_\gamma + I)}{
\det(2 P^p C_\gamma + I)^{1/2}}
\exp(d_\gamma' ((C_\gamma +(P^p)^{-1}/2)^{-1} - 
(C_\gamma + (P^p)^{-1})^{-1})d_\gamma) - 1.$$
As $\gamma \uparrow 1$, we have
\begin{equation}
\label{eq:var-weights}
N^2 \VAR{\tilde{\alpha}^{u,\gamma}_j} \sim (1-\gamma)^2 (\frac{1}{2} \textrm{tr}
(H P^p H' M)  + (y-H\mu^p)' M (y-H \mu^p)),
\end{equation} 
where
$M=(I-K_1'H')R^{-1}(I-HK_1) HP^pH' (I-K_1'H')R^{-1}(I-HK_1)$.
\end{lemma}
A proof is given in the appendix. The matrix $M$ is positive definite and
also the trace in formula (\ref{eq:var-weights}) is positive. Therefore we
expect that the variance of the weights is of the order
$\mathcal{O}(N^{-2}(1-\gamma)^2q)$: This is true if $P^p$, $R$ and $H$ are
all multiples of the identity, and there is no reason for a different
behavior in other cases.  This suggests that for high-dimensional
observations, we need to choose $\gamma$ close to one in order to avoid
degeneracy.  Note however that the final update can still differ from the
\EnKF{} update, even if the largest part of the update occurs with the
ensemble Kalman filter.

\subsection{Criteria for the selection of $\gamma$}
Because the Kalman filter update uses only the first two moments of the
forecast distribution, it seems plausible that for non-Gaussian forecast
distributions, the Kalman update will be less informative than the correct
update. Hence, as long as the spread is a meaningful measure of
uncertainty, we expect the Kalman update to have a larger spread than the
correct update. This is not always true though: The variance of the
correct posterior is only on average smaller than the 
variance $(I-K(P^p)H)P^p$ of the \EnKF{} update. In particular, this may
fail to hold for some values of $y$ if the prior is multimodal.

Still, this heuristic suggests that in some cases the spread of the update
ensemble will increase monotonically in $\gamma$ and we could choose
$\gamma$ such that the spread of the update is not smaller than a factor
$\tau$ times the spread of an \EnKF{} update (where $\tau$ is maybe between
0.5 and 0.8).  This means that we compare the variances of the Gaussian
mixture
$\sum_{j=1}^n \frac{N_j}{N} \mathcal{N}(\mu^{u,\gamma}_j, P^{u,\gamma})$,
where $N_j$ denotes the number of times component $j$ in
(\ref{update-step2}) has been selected, with the variances of the Kalman
filter update, that is the diagonal elements of $(I -
K(\hat{P}^p)H)\hat{P}^p$.  However, this is computationally demanding,
because we have to compute among other things $P^{u,\gamma}$; compare the
discussion in Section \ref{high-dim} above.

A simpler procedure is based on the standard deviations of the updated
ensembles.  Denote by $\hat{\sigma}^u_i$ the standard deviation of the
$i$-th component of the final update sample $x^u_j$ (which depends on
$\gamma$) and by $\hat{\sigma}^{u,En}_i$ the standard deviation of the
update sample using the \EnKF{}. Then we could choose the smallest $\gamma$
such that
$\sum_i \hat{\sigma}^u_i \geq \tau \sum_i \hat{\sigma}^{u,En}_{i}$
or (in order to control not only the total spread but all marginal spreads)
such that
$\sum_i \min \left(1,\hat{\sigma}^u_{i}(\hat{\sigma}^{u,En}_{i})^{-1}\right)
\geq \tau N$.
This has the advantage that it is easy to compute also in high dimensions.
The disadvantage is that it depends also on the generated noises. This can
be reduced somehow by taking the same noises $\epsilon_{1,j}$ and
$\epsilon_{2,j}$ for all values of $\gamma$ under consideration. 

A third possibility is to look only at the properties of the weights
$\alpha^{u,\gamma}_j$. We can take as the measure of the sampling diversity the
so-called effective sample size \ESS \, \citep{liu96}, which is defined as
$$\ESS = \frac{1}{\sum_j (\alpha^{u,\gamma}_j)^2} = 
\frac{N}{1 + N \sum_j (\alpha^{u,\gamma}_j-1/N)^2},$$
or the quantity
$$\DIV = \sum_{j=1}^N \min(1,N \alpha^{u,\gamma}_j) = 
N(1-\frac{1}{2} \sum_j |\alpha^{u,\gamma}_j - 1/N|),$$
which is the expected number of components that are chosen when generating
$(x^u_j)$ according to \eqref{update-step2} with balanced
sampling. Both measures are related to a distance between
the $\alpha^{u,\gamma}_j$ and uniform weights. Although there is no
universal relation between the two criteria, in typical cases 
$\ESS < \DIV$, i.e.,
$\ESS$ is a more conservative measure of diversity. Both criteria
do not take the spread in $P^{u,\gamma}$ into account, which also increases if
$\gamma$ increases. Therefore, they give only a lower bound for the
diversity, but they are easy to compute. We then choose
$\gamma$ as the smallest value for which $\ESS > \tau N$ 
or $\DIV > \tau N$. In order to avoid excessive computations,
we considered in the examples below only multiples of $1/15$ as values
for $\gamma$ and used 
a binary search tree (with at most 4 search steps), assuming that
the diversity is increasing in $\gamma$. We did not try to prove
this assumption since the calculation is expected to be extremely tedious;
the assumption is safe to make, though, since at the worst we end up with a
too large $\gamma$. Alternatively one could use an approximation
of $\ESS$ based on (\ref{eq:var-weights}). 

\section{Examples of single updates}
\label{sec:ex-single}

\subsection{Description of the setup}
We consider a single update for 4 situations. In all cases
$H=I$ and $R=\sigma^2 I$. There are 2 forecast samples
$(x^p_j)$ combined with 2 values $y=y_1$ and $y=y_2$ for the observations. 
The construction of the forecast sample starts with a sample $z_j \sim
\mathcal{N}_q(0,I)$, which is then modified to introduce non-Gaussian
features. More precisely, we consider the following situations:
\begin{itemize}
\item[1.] A Gaussian prior: We set $\sigma= 0.5$, $x^p_j=z_j$, $y_1=(0, \dots, 0)'$, 
and $y_2=(1.5,1.5,0, \dots, 0)'$. This means that the second 
observation contradicts
the prior, although not excessively.
\item[2.] A bimodal prior: We set $\sigma=3$, $x^p_j = z_j$ for $j \leq N/2$ and 
$x^p_j = z_j + (6,0, \dots,0)'$ for $j > N/2$, $y_1=(-2,0, \dots, 0)'$ and
$y_2=(3,0, \dots, 0)$. In the former case, the true posterior is
unimodal and in the latter case it is bimodal. 
\end{itemize}
We take $N=50$ and $q=10,50,250$. We generate one sample in dimension $250$
and use the first $q$ components. In all cases, we use a
triangular taper with range 10, assuming that the states are values along a
line (the optimal taper has range 0, since the true covariance $P^p$ is
diagonal).  Without a taper, all procedures break down: They become
overconfident as the dimension increases, and the estimated sampling
diversity increases with $q$.

\subsection{Variation of $\gamma$}
\begin{figure}
\centering
\includegraphics[scale=1]{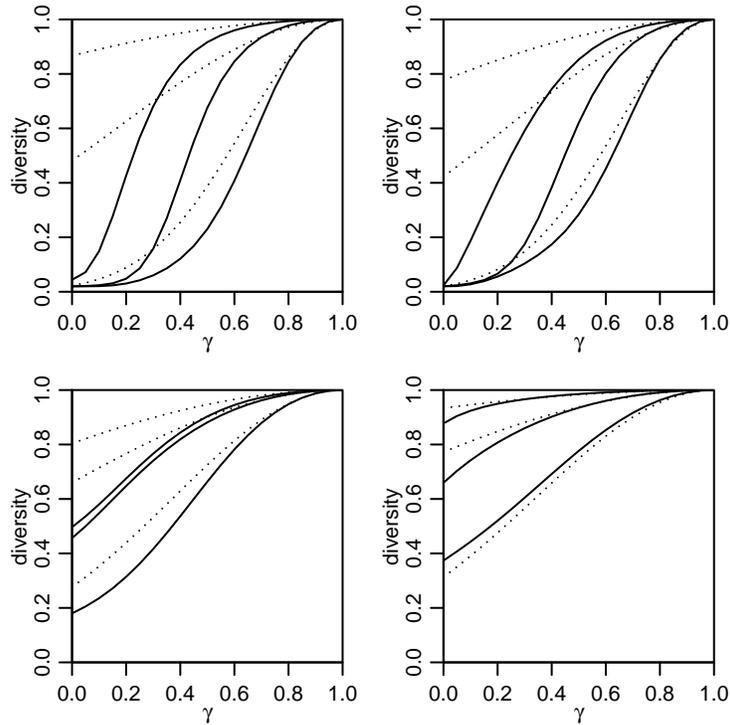}
\caption{Diversity $N^{-1} \times \ESS$ as a function of $\gamma$. Top row:
  Gaussian prior, bottom row: bimodal prior. Left column: $y_1$, right
  column $y_2$. Dimension $q$ of the state variable (from top line to
  bottom line): $q=10,50,250$. The dotted lines show the approximate
  diversity computed from \eqref{eq:var-weights} with $\mu^p$ and $P^p$
  estimated from the sample.}
\label{fig-diversity}
\end{figure}

We compute the update for $\gamma \in \{0,0.05,0.10, \dots, 1\}$ and take as
the measure of the sampling diversity the quantity $N^{-1} \times
\ESS$ introduced above. Some results are shown in
Figure~\ref{fig-diversity}. The diversity increases with $\gamma$ and
decreases with the dimension $q$ as expected.  In the bimodal case, even
the particle filter does not degenerate, and typically small or moderate
values of $\gamma$ apparently give sufficient diversity.

\subsection{The updates of the first two coordinates}
\begin{figure}
\centering
\includegraphics[scale=1]{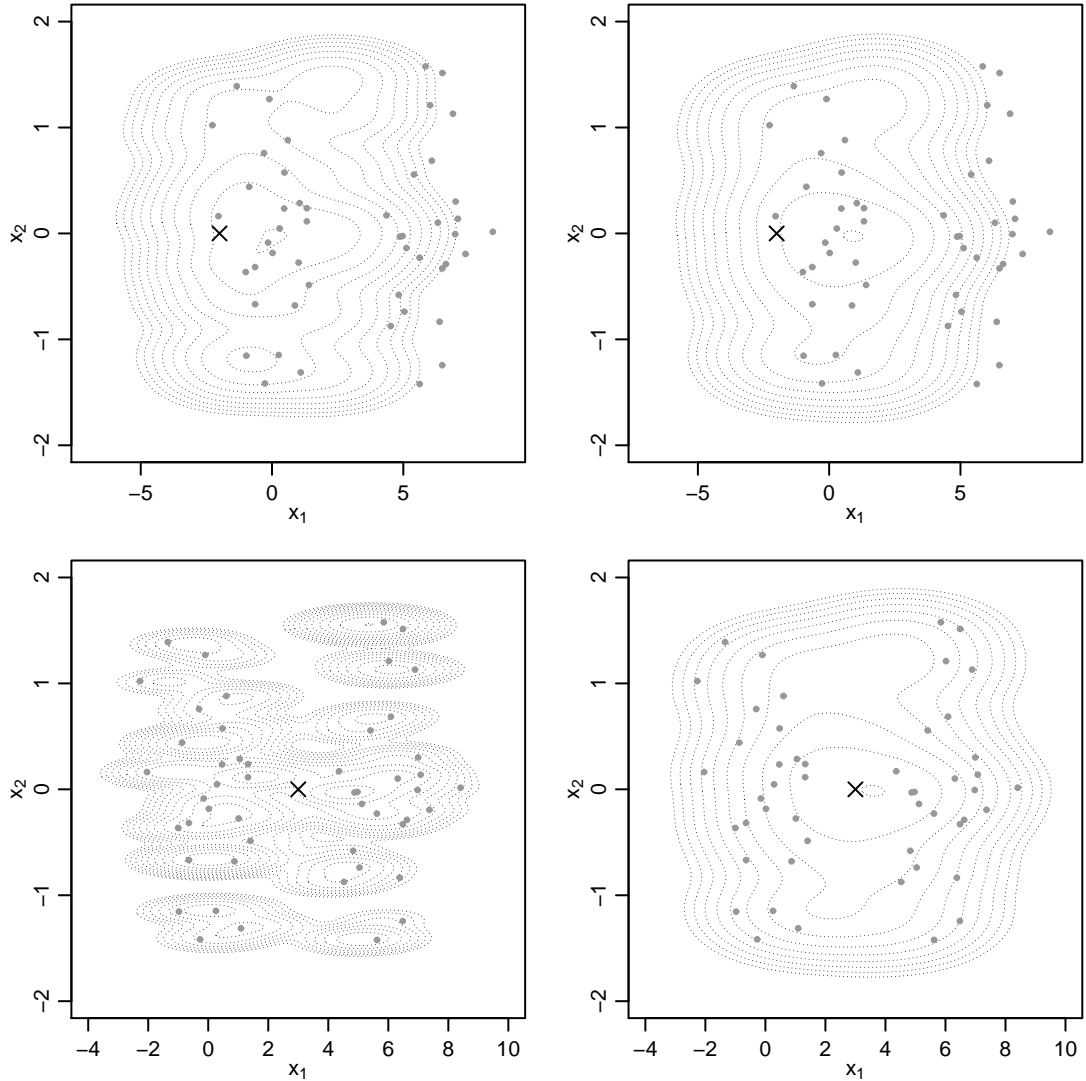}
\caption{First two components of the update of the bimodal prior with
  $q=250$.  Upper row: $y_1$, lower row: $y_2$. Left column: $\gamma$
  chosen to achieve a diversity of about 40\%. Right column: Ensemble
  Kalman filter. The prior sample is shown light grey, the observation is
  marked with a cross.  The contours show the Gaussian mixture
  \eqref{update-step2}: Levels are equidistant on a log scale such that the
  lowest level corresponds to 1\% of the maximum.}
\label{fig-update-1}
\end{figure}

We concentrate on the first two coordinates of $x^{u,\gamma}_j$ which
contain the non-Gaussian features (if present). We show the contours of the
true update density \eqref{update-step2} for the ensemble Kalman filter and
for the filter with $\gamma$ chosen such that the diversity $\tau = N^{-1}
\cdot \ESS$ is approximately 40\%. Figure~\ref{fig-update-1} shows the
results for bimodal prior with $q=250$. In case of the Gaussian prior, the
two plots (which are not shown here) are virtually identical. In the
non-Gaussian situation, the combined filter is able to pick up some
non-Gaussian features. In particular, the shape and not only the location
depends on the observation, and the bimodality of the posterior is
captured.

\section{Examples of filtering with many cycles}
\label{sec:ex-several}

\subsection{The Lorenz~96 model}
\label{sec:exL96}
The 40-variable configuration of the Lorenz~96 model \citep{lorenz96} is
governed by the ordinary differential equation
\begin{equation*}
  \frac{\mathrm{d}X^k_t}{\mathrm{d}t} = (X^{k+1}_t-X^{k-2}_t) X^{k-1}_t -
  X^k_t + 8, \quad k=1, \dots, 40.
\end{equation*}
where the boundary conditions are assumed to be cyclic, i.e.,
$X^k=X^{40+k}$. The model is chaotic and mimics the time-evolution of a
scalar meteorological quantity on a latitude circle. We adopt the same
experimental setup as in \cite{bengtsson03} and \cite{frei11}: Measurements
of odd components $X^{2k-1}$ with uncorrelated additive $\N(0,0.5)$ noise
at observation times $0.4 \times n$, $n=1,\dots,2000$, are taken. The large
lead time produces a strongly nonlinear propagation step. The system is
integrated using Euler's method with step size $0.001$. Both the \EnKF{}
and \EnKPF{} are run with $N=400$ ensemble members. The true initial state and
the initial ensemble members are randomly drawn from $\N_{40}(0,I)$. All
sample covariance matrices are replaced by tapered estimates; for the sake
of simplicity, we used the same taper matrix $C$ throughout, namely the GC
taper constructed from the correlation function given in \cite[equation
(4.10)]{gaspari99} with support half-length $c=10$.  For the \EnKPF{}, the
parameter $\gamma$ is chosen adaptively to ensure that the diversity $\tau
= N^{-1} \times \ESS$ stays within a prespecified interval
$[\tau_0,\tau_1] \subset [0,1]$ if possible.  We also ran the filter
proposed in \cite{papadakis10}. For the given ensemble size, we were not
able to obtain a non-divergent run; the filter collapsed after just a few
cycles.

The filter performance is assessed via a scoring rule evaluated at
observation times. Here, we use the root mean square error of the
ensemble mean, and the continuous ranked probability score
\citep{gneiting07} for the first two state variables. More precisely, if
$X^{k}_t$ is the true solution at time $t$, and $\widehat{X}^{k}_t$ the
mean of the updated ensemble, and $\widehat{F}_t^k(y)$ the marginal
empirical cumulative distribution function of the updated ensemble, then
the root mean square error of the ensemble mean at time $t$ is
\begin{equation}
\label{eq:RMSE}
\RMSE_t = \sqrt{\frac{1}{q} \sum_{k=1}^q
  \left(X^{k}_t-\widehat{X}^{k}_t\right)^2}
\end{equation}
and the continuous ranked probability score for the $k$th variable at
time $t$ is
\begin{equation}
\label{eq:CRPS}
\CRPS^k_t = \int_{\R} \left(\widehat{F}_t^k(y)-1_{\{y \geq
    X^k_t\}}\right)^2dy, \quad k=1,2
\end{equation}
where $t=0.4 \times n, n=1,\dots,2000$. Notice that for reasons of
symmetry, we only consider the \CRPS{} of the first two state variables
(i.e., one observed and one unobserved variable).

Tables~\ref{tab:1} and \ref{tab:2} compile summaries (first and ninth
decile, mean and median) of the 2000 \RMSE{} and \CRPS{} values.
\begin{table}
\begin{center}
\begin{tabular}{lcrrrr}
  \hline
  &$[\tau_0,\tau_1]$&$10\%$ & $50\%$ & mean & $90\%$ \\
  \hline
  EnKF&&0.56&0.81&0.87&1.25 \\
  \hline
  EnKPF&$[0.80,0.90]$&0.52&0.75&0.83&1.21 \\
  EnKPF&$[0.50,0.80]$&0.51&0.73&0.80&1.18 \\
  EnKPF&$[0.30,0.60]$&0.50&0.71&0.79&1.17 \\
  EnKPF&$[0.25,0.50]$&0.49&0.70&0.78&1.16 \\
  EnKPF&$[0.10,0.30]$&0.49&0.71&0.79&1.17 \\
  \hline
\end{tabular}
\end{center}
\caption{Lorenz 96 system, experimental setup as given in
  Section~\ref{sec:exL96}: summary statistics of
  \RMSE{} \eqref{eq:RMSE} over 2000 cycles using the \EnKF{} (EnKF)
  and the
  \EnKPF{} (EnKPF) with constrained diversity $\tau=N^{-1} \ESS \in [\tau_0,\tau_1]$ for the weights.}
\label{tab:1}
\end{table}

\begin{table}
\begin{center}
\begin{tabular}{lcrrrrrrrrr}
  \hline
  \multicolumn{2}{c}{} &
  \multicolumn{4}{c}{$X^1$} &
  \multicolumn{1}{c}{} &
  \multicolumn{4}{c}{$X^2$} \\
  \cline{3-6} \cline{8-11}
  &$[\tau_0,\tau_1]$& $10\%$ & $50\%$ & mean & $90\%$ && $10\%$ & $50\%$ & mean
  & $90\%$\\
  \hline
  EnKF&&0.12&0.22&0.32&0.65 && 0.14&0.38&0.57&1.18 \\
  \hline
  EnKPF&$[0.80,0.90]$&0.11&0.21&0.30&0.62 && 0.13&0.33&0.54&1.13 \\
  EnKPF&$[0.50,0.80]$&0.11&0.21&0.29&0.61 && 0.12&0.32&0.51&1.10 \\
  EnKPF&$[0.30,0.60]$&0.11&0.20&0.29&0.59 && 0.12&0.32&0.49&1.02 \\
  EnKPF&$[0.25,0.50]$&0.10&0.20&0.28&0.58 && 0.11&0.31&0.48&1.00 \\
  EnKPF&$[0.10,0.30]$&0.10&0.21&0.29&0.59 && 0.11&0.31&0.50&1.05 \\
  \hline
\end{tabular}
\end{center}
\caption{Lorenz 96 system, experimental setup as given in
  Section~\ref{sec:exL96}: summary statistics of \CRPS{}
  \eqref{eq:CRPS} over 2000
  cycles for the state variables $X^1$ (observed) and $X^2$ (unobserved)
  using the \EnKF{} (EnKF) and the
  \EnKPF{} (EnKPF) with constrained diversity $\tau=N^{-1} \ESS \in
  [\tau_0,\tau_1]$ for the weights.} 
\label{tab:2}
\end{table}
The gain over the \EnKF{} achieved by the \EnKPF{} is substantial. In
particular, as the \CRPS{} values show, the \EnKPF{} is able to track the
unobserved states much more accurately than the \EnKF{}. Overall, the
results for the \EnKPF{} are comparable with those reported for the XEnKF
in \cite{frei11}. Arguably, the best performance of the \EnKPF{} is
achieved with diversity constrained to $[0.25,0.50]$, but the scores are
surprisingly robust. Finally, we note that for smaller ensemble sizes,
e.g., $N=100$, the \EnKPF{} still improves over the \EnKF{}, but the
results (which are not shown here) are less impressive. Nevertheless, one
should keep in mind that with a particle filter, even with judicious
tuning, much more particles are required to compete with the \EnKF{}, as
illustrated in \cite{bocquet10} (for a slightly different configuration of
the Lorenz model).

\subsection{The Korteweg-de Vries equation}
\begin{figure}
\centering
\includegraphics[scale=1]{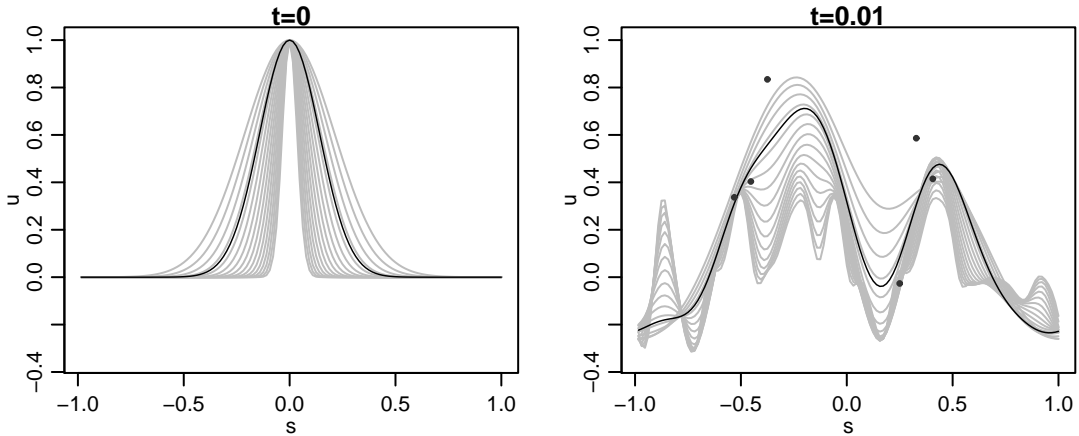}
\caption{Korteweg-de Vries equation. The left figure shows the initial
  16-member ensemble (grey) at time $t=0$ together with the true solution
  (black). The right figure shows the predictive ensemble (grey) at the
  first observation time $t=0.01$ together with the observations (black
  bullets) and the true solution (black).}
\label{fig:KdV-illustration}
\end{figure}

We consider the Korteweg-de Vries equation on the circle
\citep{drazin89}:
\begin{equation*}
\partial_t x + \partial^3_s x + 3 \partial_s x^2 = 0
\end{equation*}
with domain $(s,t) \in [-1,1) \times [0,\infty)$ and periodic boundary
conditions, $x(s=-1,t)=x(s=1,t)$. Versions of
this equation have been used as test beds for data assimilation in, e.g.,
\cite{leeuwen03}, \cite{lawson05}, or \cite{zupanski06}. The spatial domain
$[-1,1)$ is discretized using an equispaced grid with $q=128$ grid
points. The spectral split step method is used to solve the equation
numerically (with an explicit 4th order Runge-Kutta time step for the
nonlinear part of the equation). As initial prior we take the random field
\[ X(s,t=0) = \exp\left(-\frac{s^2}{\eta^2}\right), \quad \log(\eta) \sim
\mathcal{U}(\log(0.05),\log(0.3)). \] For the truth, we use $\eta=0.2$, and
the initial ensemble is a (quasi-random) sample from $X(s,t=0)$. The
ensemble size is $N=16$, and thus $N \ll q$. Six irregularly spaced
observations with uncorrelated additive $\N(0,0.02)$ noise at observation
times $0.01 \times n$, $n=1,\dots,10$, are taken. For illustration,
Figure~\ref{fig:KdV-illustration} displays the initial 16-member ensemble
and the predictive ensemble at the first observation time (together with
the observations).

\begin{figure}
\centering
\includegraphics[scale=1]{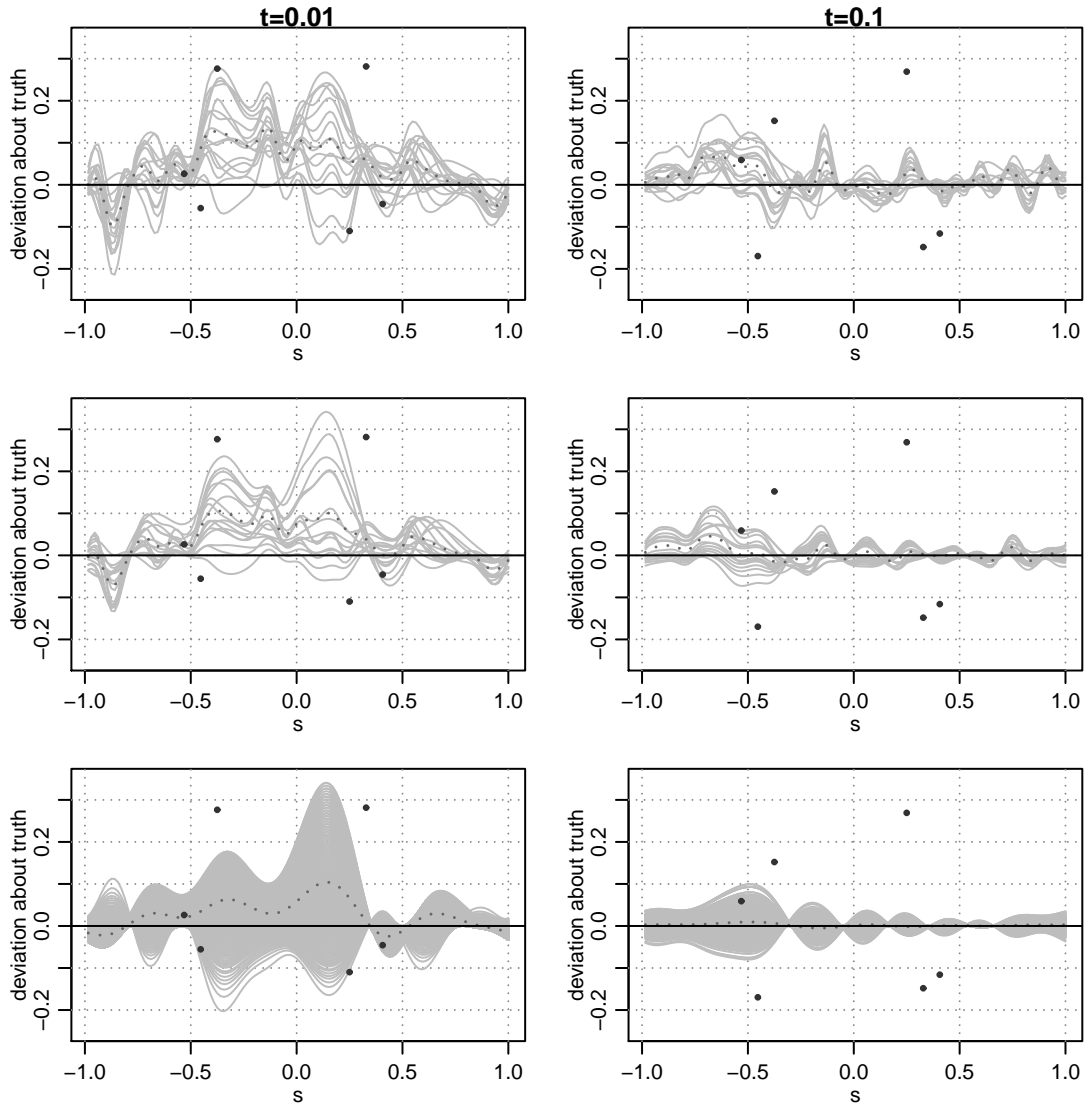}
\caption{Korteweg-de Vries equation. Ensemble deviations about the truth,
  i.e., filtered ensemble minus true solution, after the 1st ($t=0.01$,
  left panel) and 10th ($t=0.1$, right panel) update cycle, for the \EnKF{}
  and \EnKPF{} with $N=16$ particles (top two rows), and for the particle
  filter with $N=256$ particles (bottom row). The solid grey lines are the
  deviations, the dotted grey lines the average of the deviations, and the
  black bullets are the observations minus the truth.}
\label{fig:KdV-deviations}
\end{figure}

\begin{figure}
\centering
\includegraphics[scale=1]{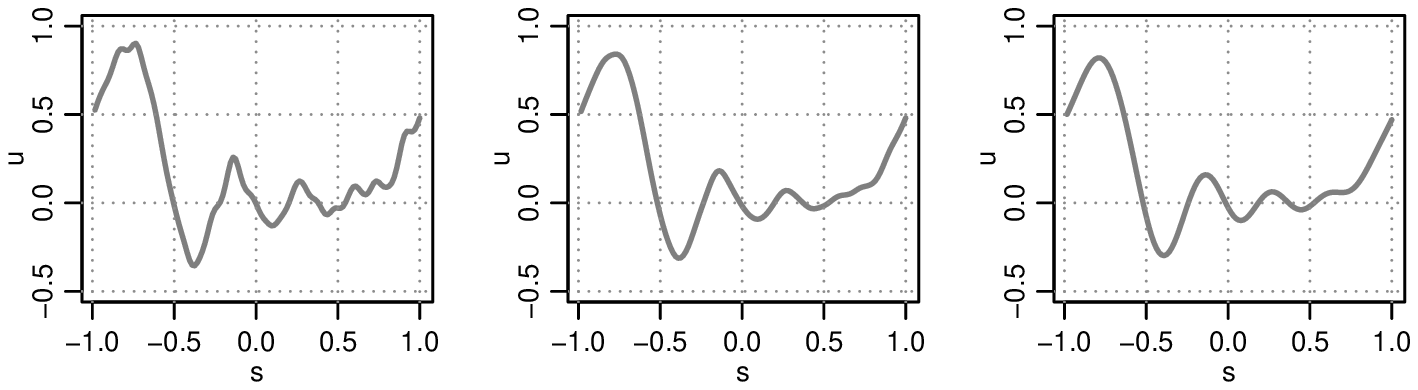}
\caption{Korteweg-de Vries equation. Particle with most curvature after 10
  update cycles ($t=0.1$), for the \EnKF{} (left), \EnKPF{} (middle) and
  particle filter (right).}
\label{fig:KdV-curvature}
\end{figure}

The particle filter, the \EnKF{} and the \EnKPF{} are run (with no tapering
applied). For the \EnKPF{}, we fix $\gamma=0.05$, which ensures that $\tau =
N^{-1} \cdot \ESS$ lies roughly in the interval $[0.80,0.90]$. Since the
particle filter degenerates very quickly for such a small ensemble, a
benchmark run with $N=256$ particles is carried
out. Figure~\ref{fig:KdV-deviations} displays ensemble deviations from the
true solution after 1 and 10 update cycles.
Apparently, both the \EnKF{} and \EnKPF{} track the true solution reasonably
well, and the state uncertainty is well represented. In terms of error of
the ensemble mean, there is not much difference between the \EnKF{} and the
\EnKPF{}. However, the \EnKPF{} produces particles that exhibit less dynamical
inconsistencies. In Figure~\ref{fig:KdV-curvature}, for each filter, the
particle with the most curvature after 10 update cycles is shown, where the
curvature of a solution $x(s,t)$ is defined by $\int_{-1}^{1} |\partial^2_s
x| (1+(\partial_s x)^2)^{-3/2} \mathrm{d}s$, and a finite difference
approximation is used for the discretized solutions. For reference, we
note that the true solution (not shown in the plots) is virtually identical
to the particle shown in the rightmost plot.
The particle filter (which conserves dynamical constraints) yields very smooth
particles, whereas the \EnKF{} may produce wiggly, ``unphysical''
particles. The \EnKPF{} lies in-between: some particles still show slight
dynamical imbalances, but these are much less pronounced than for the
\EnKF{}. Also, if a taper is applied to the forecast covariance matrix
(which is not the case in the example here), the
\EnKF{} suffers even more from these imbalances.

\section{Outlook to possible generalizations}
\label{sec:outlook}

In the spirit of the ``progressive correction'' idea \citep{musso01}, the
\EnKPF{} update could also be split up in several steps. We fix constants
$\gamma_i>0$ and $\delta_i>0$ with $\sum_{i=1}^L \gamma_i + \delta_i =
1$. Then, for $i=1,\dots,L$, we apply an \EnKF{} update with likelihood
$\ell(x|y)^{\gamma_i}$, followed by a particle filter update with
likelihood $\ell(x|y)^{\delta_i}$, followed by the resampling step. It is
only necessary to estimate the predictive covariance for the first step;
for the subsequent steps, $i=2,\dots,L$, we can compute the covariance
analytically from the mixture representation \eqref{update-step2} (for
large $q$, this is numerically delicate, but the same remedies as discussed
in Section~\ref{high-dim} can be applied). We expect that the bias of such
an iterative \EnKPF{} update is similar as for a single \EnKPF{} update with
$\gamma=\sum_{i=1}^L \gamma_i$, but the variance will decrease with
increasing $L$ since the likelihoods become flatter. In the limiting case
$\sum_{i=1}^L \gamma_i \to 0$, which corresponds to a full particle filter
update, we conjecture that $L=\mathcal{O}(q)$ is sufficient to retain the
sampling diversity. This claim is supported by \cite{beskos12}, who analyze
the ``tempering'' idea in simpler but related situations.

A potential drawback of the \EnKPF{} (in comparison to non-Gaussian ensemble
filters akin to \cite{lei11}) is its restriction to Gaussian linear
observations. However, the idea of combining an ensemble Kalman filter and
a particle filter update could also be used for arbitrary observation
densities. Let $H$ be a matrix that selects those components of the state
variable that influence the observation, and assume that we have an
approximation of the likelihood of the form $\ell(Hx|y) \approx
\varphi(g(y);Hx, R(y))$. Then we can use this approximation for an \EnKF{}
update, and correct by a particle filter update with weights proportional
to $\ell(Hx^u_j|y) (\varphi(g(y);Hx^u_j, R(y)))^{-1}$. In order to
construct an approximation of the likelihood of the above form, we can use
a Taylor approximation
\begin{equation*}
  \log \ell(Hx|y) \approx \log \ell(H\mu^p|y) + a(y)'H(x-\mu^p) \\
  + \frac{1}{2} (x-\mu^p)'H'b(y)H(x-\mu^p)
\end{equation*}
where $a(y)$ and $b(y)$ are the gradient and the Hessian, respectively, of
the log likelihood. Then $R(y)=-b(y)^{-1}$ and $g(y)=R(y)a(y)$.
Alternatively, one could center the expansion at the mode of
likelihood. Such an approximation is expected to work well in cases where
the likelihood is log-concave, e.g., when $y$ given $x$ is Poisson with
parameter $\exp(x)$.

\section*{Acknowledgments} The authors thank Jo~Eidsvik for fruitful
discussions.

\appendix
\section*{Appendix}

\subsection*{Proof of Theorem~\ref{thm:consistency}}
Convergence of $\pi^{u,N}_{\rm{EnKPF}}$ to $\pi^u$ implies convergence of
$\frac{1}{N}\sum_{j=1}^N \Delta_{x^{u,N}_j}$ to $\pi^u$, see Lemma~7 in
\cite{frei11}. It remains to establish the former convergence. To begin
with, we introduce some notation used throughout the remainder. For ease of
legibility, the dependence on $N$ is dropped. An overbar $\overline{\cdot}$
is added to any quantity to denote its population counterpart, in which
$\widehat{P}^p$ has been replaced by $P^p$. For a measure $\pi$ and a
function $h$, we write $\pi h=\int h(x) \pi(dx)$. Expressions of the form
$A \to B$ are shorthand for $A$ converges to $B$ almost surely as $N$ goes
to $\infty$. Straightforward application of the strong law of large numbers
shows that the population version $\overline{\pi}^{u,N}_{\rm{EnKPF}}$
converges to some nonrandom limit, and it is clear by construction that
this limit equals $\pi^u$. Hence, it remains to prove that the population
version of the \EnKPF{} has the same limit as the \EnKPF{}, i.e., we need
to prove that
\begin{equation}
\label{eq:pih-overbarpih}
|\pi^u_\mathrm{EnKPF}h - \overline{\pi}^u_\mathrm{EnKPF} h| \to 0
\end{equation}
for any continuous and bounded function $h$. In addition, we may assume
that $h$ is compactly supported on, say, $S_h \subset \R^q$, since these
functions are still convergence determining for the weak topology. Write
$||h||_\infty = \max_x |h(x)|$.

Observe that
\begin{equation}
\label{eq:cons-1and2}
\begin{split}
  |\pi^u_\mathrm{EnKPF}h - \overline{\pi}^u_\mathrm{EnKPF} h| &\leq
  \left|\sum_{j=1}^N \alpha^u_j \N(\mu^u_j,P^u)h
    - \sum_{j=1}^N \overline{\alpha}^u_j \N(\mu^u_j,P^u)h \right| \\
  &\quad + \left|\sum_{j=1}^N \overline{\alpha}^u_j \N(\mu^u_j,P^u)h -
    \sum_{j=1}^N \overline{\alpha}^u_j
    \N(\overline{\mu}^u_j,\overline{P}^u)h \right|.
\end{split}
\end{equation}
In the following, we show that both terms on the right-hand side of
\eqref{eq:cons-1and2} converge to $0$, which proves
\eqref{eq:pih-overbarpih}. For later use, we note that $\widehat{P}^p \to
P^p$, and hence by continuity, 
\begin{equation}
\label{eq:cons-moments}
K(\gamma \widehat{P}^p)  \to K(\gamma P^p), \quad
Q(\gamma,\widehat{P}^p)  \to Q(\gamma,P^p), \quad
K((1-\gamma)Q(\gamma,\widehat{P}^p)) \to K((1-\gamma)Q(\gamma,P^p)), \quad
P^u \to \overline{P}^u.
\end{equation}

The first term in \eqref{eq:cons-1and2} can be bounded by $||h||_\infty
\sum_{j=1}^N |\alpha^u_j-\overline{\alpha}^u_j|$.  Let $w^u_j$ and
$\overline{w}^u_j$ denote the unnormalized weights in \eqref{eq:alpha},
i.e.,
\[ w^u_j=\varphi(y;H \nu^{u,\gamma}_j,H Q(\gamma,\hat{P}^p)H' +
\frac{1}{1-\gamma}R). \]
Observe that
\begin{equation*}
w^u_j
\leq \varphi(0;0,HQ(\gamma,\widehat{P}^p)H' + \frac{1}{1-\gamma}R)
\leq \varphi(0;0,\frac{1}{1-\gamma}R),
\end{equation*}
where the last inequality follows from $\mathrm{det}(M+N) \geq
\mathrm{det}(M)$ for arbitrary positive definite $M$ and positive
semi-definite $N$. The same bound holds true for $\overline{w}^u_j$.
Notice that
\begin{equation*}
\begin{split}
  \sum_{j=1}^N \left|\alpha^u_j-\overline{\alpha}^u_j\right| \leq
  \frac{1}{\mathrm{ave} \, w^u} \frac{1}{N} \sum_{j=1}^N
  \left| w^u_j-\overline{w}^u_j \right| + \frac{1}{N} \sum_{j=1}^N
  \overline{w}^u_j \left|\frac{1}{\mathrm{ave} \, w^u} -
  \frac{1}{\mathrm{ave} \, \overline{w}^u} \right|
\end{split}
\end{equation*}
and
$\left|\mathrm{ave} \, w^u - \mathrm{ave} \, \overline{w}^u \right|
\leq \frac{1}{N} \sum_{j=1}^N \left|w^u_j-\overline{w}^u_j\right|$,
where $\mathrm{ave} \, w^u = \frac{1}{N} \sum_{j=1}^N w^u_j$ and
$\mathrm{ave} \, \overline{w}^u = \frac{1}{N} \sum_{j=1}^N
\overline{w}^u_j$. The $\overline{w}^u_j$ are iid, hence, $\mathrm{ave} \,
\overline{w}^u$ converges almost surely, and we conclude that
$\sum_{j=1}^N |\alpha^u_j-\overline{\alpha}^u_j| \to 0$ if
\begin{equation}
\label{eq:cons-1}
\frac{1}{N} \sum_{j=1}^N \left|w^u_j-\overline{w}^u_j\right| \to 0.
\end{equation}
To show \eqref{eq:cons-1}, we fix a compact set $D \subset \R^q$. Then
we have
\begin{equation}
\begin{split}
  \frac{1}{N} \sum_{j=1}^N \left|w^u_j-\overline{w}^u_j\right| &\leq
  \frac{1}{N} \sum_{j=1}^N \left|w^u_j-\overline{w}^u_j\right| 1_{x^p_j
    \notin D} + \max_{1 \leq j \leq N : x^p_j \in D}
  \left|w^u_j-\overline{w}^u_j\right| \\
  &\leq 2 \varphi(0;0,\frac{1}{1-\gamma}R) \frac{1}{N} \sum_{j=1}^N
  1_{x^p_j \notin D} + \max_{1 \leq j \leq N : x^p_j \in D}
  \left|w^u_j-\overline{w}^u_j\right| \\
  &\to 2 \varphi(0;0,\frac{1}{1-\gamma}R) \cdot \PR{x^p_1 \notin D},
\end{split}
\end{equation}
where the ``max''-term goes to zero for reasons of uniform continuity in
combination with \eqref{eq:cons-moments}. Letting $D \uparrow \R^q$,
establishes \eqref{eq:cons-1} by virtue of dominated convergence.

We now analyze the second term in \eqref{eq:cons-1and2}.  Again, we fix a
compact set $D \subset \R^q$. Then we have
\begin{equation*}
\begin{split}
\sum_{j=1}^N \overline{\alpha}^u_j \left| \N(\mu^u_j,P^u)h -
    \N(\overline{\mu}^u_j,\overline{P}^u)h \right|
&\leq \max_{1 \leq j \leq N: x^p_j \in D} ||h||_\infty
   \int_{S_h} \left|\varphi(z;\mu^u_j,P^u)
               -\varphi(z;\overline{\mu}^u_j,\overline{P}^u)\right|dz \\
& \qquad + 2 ||h||_\infty \sum_{j=1}^N \overline{\alpha}^u_j 
               1_{x^p_j \notin D} \\
& \to 2 ||h||_\infty \, \frac{\ERW{\overline{w}^u_1
    1_{x^p_1 \notin D}}}{\ERW{\overline{w}^u_1}},
\end{split}
\end{equation*}
where the ``max''-term goes to zero for reasons of uniform continuity in
combination with \eqref{eq:cons-moments}. Letting $D \uparrow \R^q$, shows
that also the second term in \eqref{eq:cons-1and2} converges to $0$, which
completes the proof.

\subsection*{Proof of Lemma~\ref{lemma:weights}}
We set 
$$Z=\exp(- \frac{1}{2}(x^p_j -\mu^p)' C_\gamma (x^p_j -\mu^p) + 
d_\gamma' (x^p_j -\mu^p)).$$
Then by definition
$$\VAR{\tilde{\alpha}^{u,\gamma}_j} = \frac{1}{N^2} \left(
\frac{\ERW{Z^2}}{\ERW{Z}^2} - 1 \right).$$ 
The lemma follows by completing the square and using that Gaussian
densities integrate to one. More precisely, for any $x$ and any 
positive definite matrix $\Gamma$: 
$$x' (C_\gamma + \Gamma) x - 2 d_\gamma' x = (x-(C_\gamma +\Gamma)^{-1}d_\gamma)' 
(C_\gamma +\Gamma) (x-(C_\gamma +\Gamma)^{-1}d_\gamma) - d_\gamma'
(C_\gamma +\Gamma)^{-1} d_\gamma.$$ 
Therefore
$$\ERW{Z} =\left(
\det(P^p) \det(C_\gamma + (P^p)^{-1})\right)^{-1/2} \exp(\frac{1}{2}
d_\gamma'(C_\gamma+ (P^p)^{-1})^{-1} d_\gamma)$$
and 
$$\ERW{Z^2}=\left(
\det(P^p)\det(2 C_\gamma + (P^p)^{-1})\right)^{-1/2} \exp(2
d_\gamma'(2C_\gamma+(P^p)^{-1})^{-1}d_\gamma).$$ 
Taking these results together, the first claim follows.

For the second claim, we note that as $\gamma \uparrow 1$
\begin{align*}
  C_\gamma &\sim (1-\gamma) H'(I-K_1'H') R)^{-1} 
(I-H K_1) H,\\
 d_\gamma &\sim (1-\gamma) H'(I-K_1'H') R)^{-1} (I-H K_1)(y-H \mu^p)
\end{align*}
because $K_\gamma$ and $Q_\gamma$ are continuous. The result then follows
by a straightforward computation. 

\bibliographystyle{apalike}
\bibliography{literature}
\end{document}